\documentclass[journal=jctcce,manuscript=article]{achemso}
\usepackage{graphicx}

\title{Probing the Statistical Validity of the Ductile-to-Brittle Transition in Metallic Nanowires using GPU Computing}

\author{William R. French}
\affiliation[Vanderbilt University]{Department of Chemical and Biomolecular Engineering, Vanderbilt University, Nashville, TN}
\author{Amulya K. Pervaje}
\affiliation[Vanderbilt University]{Department of Chemical and Biomolecular Engineering, Vanderbilt University, Nashville, TN}
\author{Andrew P. Santos}
\affiliation[North Carolina State University]{Department of Chemical and Biomolecular Engineering, North Carolina State University, Raleigh, NC}
\author{Christopher R. Iacovella}
\affiliation[Vanderbilt University]{Department of Chemical and Biomolecular Engineering, Vanderbilt University, Nashville, TN}
\author{Peter T. Cummings}
\email{peter.cummings@vanderbilt.edu}
\affiliation[Vanderbilt University]
{Department of Chemical and Biomolecular Engineering, Vanderbilt University, Nashville, TN}
\alsoaffiliation[Oak Ridge National Laboratory]{Center for Nanophase Materials Sciences, Oak Ridge National Laboratory, Oak Ridge, TN}

\begin{document}

\pagebreak

\begin{abstract}
 We perform a large-scale statistical analysis ($>$ 2000 independent simulations) of the elongation and rupture of gold nanowires, probing the validity and scope of the recently proposed ductile-to-brittle transition that occurs with increasing nanowire length [Wu $et.$ $al.$, Nano Lett. {\bf12}, 910-914 (2012)].  To facilitate a high-throughput simulation approach, we implement the second-moment approximation to the tight-binding (TB-SMA) potential within HOOMD-Blue, a molecular dynamics package which runs on massively parallel graphics processing units (GPUs). In a statistical sense, we find that the nanowires obey the ductile-to-brittle model quite well; however, we observe  several unexpected features from the simulations that build on our understanding of the ductile-to-brittle transition. First, occasional failure behavior is observed that qualitatively differs from that predicted by the model prediction; this is attributed to stochastic thermal motion of the Au atoms and occurs at temperatures as low as 10 K. In addition, we also find that the ductile-to-brittle model, which was developed using classical dislocation theory, holds for nanowires as small as 3 nm in diameter. Finally, we demonstrate that the nanowire critical length is higher at 298 K relative to 10 K, a result that is not predicted by the ductile-to-brittle model. These results offer practical design strategies for adjusting nanowire failure and structure, and also demonstrate that GPU computing is an excellent tool for studies requiring a large number independent trajectories in order to fully characterize a system's behavior.
\end{abstract}

\textsc{}


\vspace{0.5in}

\section{Introduction}

Understanding the rupture process of elongating metallic nanowires (NWs) \cite{Richter:2009,Seo:2011,Yue:2011,Yue:2012} under a range of conditions is important in areas such as nanoelectronics \cite{Nitzan:2003,Vazquez:2012,French:2012,French-fluctuations:2013,French-macs:2013,Dubi:2013} and nanoscale cold welding \cite{Lu:2010}, where the structure of a NW may affect its properties. For example, the deformation of a NW can significantly alter the electron transport properties of atomic-\cite{Iacovella:2011} and molecular-scale\cite{French-macs:2013,French-fluctuations:2013} junctions. Recently, Wu and co-workers \cite{Wu:2012} suggested a transition from ductile-to-brittle failure for mechanically deformed NWs as the NW length is increased.  The ductile regime, where virtually all previous simulation studies \cite{French:2012,French:2011,Iacovella:2011,Iacovella_scidac:2011,Pu_JCP:2007,Pu:2008,Garcia-Mochales:2013,Gall:2004,Coura:2004,Sato:2005,Wang:2007,Koh:2005,Koh:2006,Park:2005} have focused, exhibits a diverse set of structural evolution modes, which, while important for producing novel nanoscale structures such as monatomic chains \cite{Coura:2004,Sato:2005,Koh:2005,Koh:2006,French:2011,Yanson:1998,Scheer:1998}, helices \cite{Koh:2005,Park:2005,French:2011}, and polytetrahedra \cite{Iacovella:2011,Garcia-Mochales:2013}, may be undesirable in applications requiring high reproducibility. In contrast, brittle failure is characterized by sudden shearing along a single slip plane, resulting in greater structural consistency. This consistency may be useful in experiments of molecular junctions, as molecules are often bridged across the tips of a broken NW, \cite{Reed:1997} and it has been established \cite{Muller:2006,Haiss:2009,French:2012,French-fluctuations:2013,French-macs:2013} that tip structure strongly influences the transport properties of the bridged molecule.  Thus, adjusting the length of a NW may provide a method for controlling the structure and properties of molecular electronic devices.

In order to facilitate reproducibility and improved control for device applications, it is important to understand the validity and scope of the ductile-to-brittle transition under a range of conditions. In their study, Wu $et$ $al.$ \cite{Wu:2012} focused on very large (from a computational cost standpoint) NWs, with diameters of 20 nm and lengths spanning from 188 to 1503 nm. Different breaking behavior may occur for significantly thinner NWs, such as the 1.8-nm core-shell \cite{Oshima:2003} or 3-nm single crystalline \cite{Lu:2010} Au NWs fabricated in experiment. In these ultra-thin NWs the impact of surface energy is more prominent, stochastic atomic motion may play an increased role, and classical dislocation plasticity may no longer apply. It also remains unclear what role temperature plays in the length-dependent mechanism. Temperature has been shown\cite{Koh:2005} to influence the mode of NW failure as well as the structural evolution of ductile nanowires\cite{Pu:2008,Iacovella_scidac:2011}. Moreover, due to the high computational cost of their simulations, Wu $et$ $al.$ \cite{Wu:2012} were limited to a single run for each NW size, and thus their results may not be representative of typical behavior since NW elongation and rupture is a stochastic process,\cite{Pu:2008,French:2011,Iacovella:2011} especially within the ductile regime. Dislocation events occurring in response to mechanical loading are highly sensitive to the relative positions of metal atoms; thus, slight differences in atoms' relative positions induced by thermal motion can cause vastly different structural pathways for two independent runs of a NW elongated under identical conditions. A follow-up study considering a large number of independent trajectories for each state point would provide statistical insight into the NW elongation process and be valuable for clarifying the validity and scope of the ductile-to-brittle transition.

Running molecular dynamics (MD) simulations on graphics processing units (GPUs)\cite{Anderson:2008,Stone:2010,Brown:2011,Brown:2012} provides an efficient means for running a large number of replicates in order to better describe the statistical behavior of NW rupture. HOOMD-Blue is a MD package built from the ground up with GPU computing in mind, and large performance boosts have been achieved with HOOMD-Blue relative to CPU-based simulations. \cite{Anderson:2008} Since it is optimized for single GPU simulations, HOOMD-Blue provides an ideal platform for high-throughput studies, such as the one we aim to conduct here. Early development of HOOMD-Blue has emphasized basic MD functionality and interaction models focusing on soft-matter systems.\cite{Anderson:2008,Nguyen:2011,Phillips:2011,LeBard:2012,Levine:2011}   Features that enable the simulation of hard-matter systems have also been added, such as the embedded-atom method (EAM) \cite{Morozov:2011}. EAM is a many-body potential designed to capture metallic bonding interactions, with resulting performance gains on par with a pairwise potential. However, prior work \cite{Pu_JCP:2007} has shown that the EAM potential overestimates the surface energy in elongating NWs, resulting in energetic and structural evolution that does not match quantum mechanical calculations. The second-moment approximation to the tight-binding (TB-SMA) potential is better suited for describing NW elongation. \cite{Pu_JCP:2007} Here, we report the implementation of the TB-SMA potential into HOOMD-Blue, provide benchmarks compared to our prior CPU implementation, and apply it to probe the ductile-to-brittle transition proposed by Wu and co-workers.

\section{Methods}

\subsection{TB-SMA potential}

Simple pairwise potentials such as the 12-6 Lennard-Jones potential fail to properly describe many of the properties ($e.g.$, vacancy formation energies, surface structure, and relaxation properties) of transition metals. \cite{Cleri:1993} Semi-empirical potentials, whose functional forms are derived from electronic structure considerations and then fit to experimental data,\cite{Daw:1984} are better suited for simulations of transition metals. For instance, TB-SMA \cite{Cleri:1993} contains a many-body term that is modeled after the square-root dependence of the band energy on the second moment electron density of state:

\begin{equation}E_{B}^{i}=-\left\{\sum_{j}\xi^{2}e^{-2q(r_{ij}/r_{0}-1)}\right\}^{1/2},\label{manybodyterm}
\end{equation} where $E_{B}^{i}$ is the many-body energy of atom $i$. TB-SMA also contains a pairwise repulsive term given by

\begin{equation}E_{R}^{i}=\sum_{j}Ae^{-p(r_{ij}/r_{0}-1)}.\label{pairwiseterm}
\end{equation} The total TB-SMA energy is then

\begin{equation}E_{C}=\sum_{i}(E_{B}^{i}+E_{R}^{i}).\label{totalenergy}
\end{equation} Values for the parameters $A$, $\xi$, $p$, $q$, and $r_{0}$ for Au are shown in Table 1, and were obtained from fits to the Au experimental cohesive energy, lattice parameter, and elastic constant. \cite{Cleri:1993} We apply an energy cutoff, $r_{cut}$, of 5.8 \AA, such that any pair of atoms separated by a distance greater than $r_{cut}$ do not interact. Differentiating eq 1 yields an expression for force that depends on the electron density, $\rho$, of atoms $i$ and $j$, where $\rho_{i}$ = $\sum_{j}e^{-2q(r_{ij}/r_{0}-1)}$. 

We port TB-SMA into HOOMD-Blue using an implementation that is similar to the EAM implementation of Morozov and co-workers: \cite{Morozov:2011} the total force acting on each atom is calculated in two stages ($i.e.$, two CUDA kernels), with each stage looping over atom $i$'s neighbors. This amounts to an additional computational cost compared to pairwise interaction models, where only a single loop over atom $i$'s neighbors is needed. Moreover, while computing the TB-SMA force, each block of threads on the GPU must be synchronized after computing $\rho$ in the first loop; this additional communication step also reduces performance. For example, for a wire containing $\sim$53,000 Au atoms there is a 35\% increase in run time due to thread synchronization for 100,000 time steps performed with fixed particle positions. Nevertheless, in ref.  \cite{Morozov:2011} the speedups of the GPU code over a CPU implementation are similar to those obtained for a simple Lennard-Jones (LJ) interaction (where thread synchronization is not needed). This suggests that the large number of arithmetic operations required by EAM and TB-SMA enables the GPU code to compensate for the performance penalty incurred by thread synchronization, resulting in an overall speedup that is comparable to a pairwise potential.

%
%
\begin{table}[] 
\caption{TB-SMA parameters for Au. \label{table:params}}
\begin{tabular}{c c c c c}
\hline
\hline
$A$ (eV) & $\xi$ (eV) & $p$ & $q$ & $r_{0}$ (\AA) \\
\hline
0.2061 & 1.790 & 10.229 & 4.036 & 4.079  \\
\hline
\hline
\end{tabular}
\end{table}

\subsection{Simulation details}

We follow the simulation protocol of our previous work \cite{Pu_JCP:2007,Pu:2008,French:2011,Iacovella:2011} to simulate NW elongation. Two rigid layers of ``gripping'' atoms (colored in green and red in Figure 1) are placed on both ends of a [100]-oriented, cylindrical NW. The gripping atoms on the left and right sides of the wire are periodically displaced by 0.05 \AA\ in the [$\bar{1}$00] and [100] directions, respectively, between 20 ps of MD. The temperature is controlled with the Nos\'{e}-Hoover thermostat and the equations of motion are integrated using the velocity Verlet algorithm with a time step of 2.0 fs. The NWs vary in their initial diameter, $D_{0}$, from 3.1-6.0 nm, while the initial length, $L_{0}$, is varied between 20-140 nm. The smallest NW is shown in Figure 1. Note that a small, ring-shaped notch is introduced in the center of the NW to control the break location. Each independent elongation simulation is initialized with a random Gaussian distribution of atomic velocities resulting in a temperature of 0.01 K. Prior to simulating elongation, we equilibrate the NW using the following method: (1) 100 ps of MD with a target pressure of zero (applied with a Nos\'{e}-Hoover barostat) in the [100] direction and a target temperature of 0.01 K; (2) 400 ps of MD, ramping the temperature from 0.01 K to the target value; (3) 400 ps of MD at the target temperature.

%
%
\begin{figure}[]
       \centering
	\includegraphics[width=4.0in]{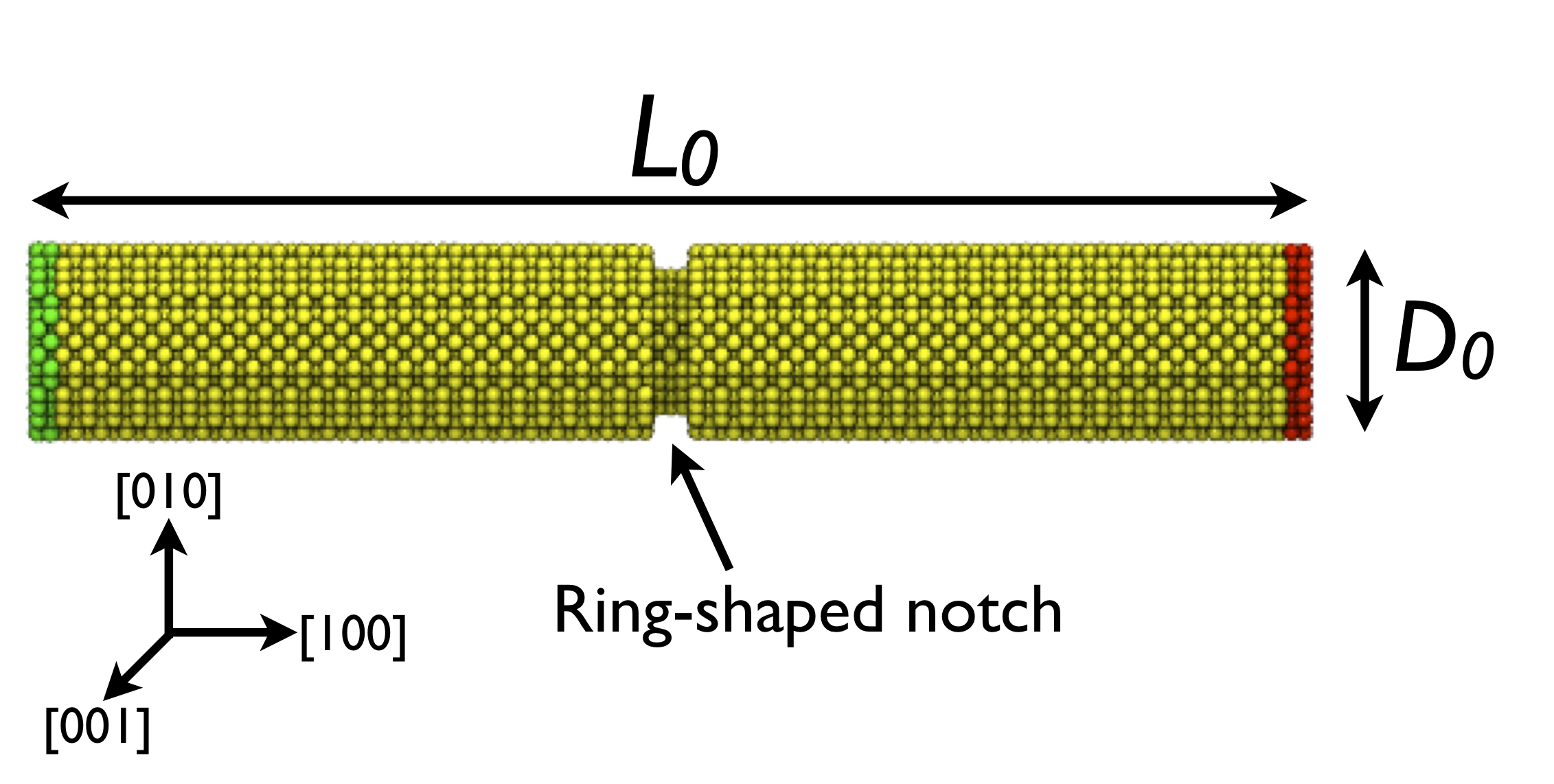}
	\caption{Schematic of an initial NW geometry. In this case, $D_{0}$ = 3.1 nm and $L_{0}$ = 20.4 nm. The ring-shaped notch is three atoms wide and two atoms deep. The gripping atoms are colored in green and red, while dynamic atoms are colored yellow.\label{fig:nanowire} }
\end{figure}   

We analyze NW trajectories by calculating the stress-strain relationship. The engineering strain, $\epsilon$, is calculated using the expression

\begin{equation}
\epsilon = \frac{L - L_{0}}{L_{0}},\label{strain}
\end{equation} 

\noindent where $L$ is the instantaneous length of the wire. The stress, $\sigma_{xx}$, along the direction of stretching is calculated with the virial expression \cite{Zimmerman:2004,Subramaniyan:2008}

\begin{equation}\sigma_{xx} =\frac{1}{V}\sum_{i}\Big[\frac{1}{2}\sum_{j}r_{x,ij}F_{x,ij}-mv_{x,i}v_{x,j}\Big],\label{stress}
\end{equation} where $V$ is the volume of the nanowire, $r_{x,ij}$ and $F_{x,ij}$ are the inter-atomic distance and the force between atoms $i$ and $j$ in the $x$ direction, $m$ is the mass of a Au atom, and $v_{x,i}$ ($v_{x,j}$) is the velocity of atom $i$ ($j$) in the $x$ direction. In accordance with previous work \cite{Koh:2005}, $V$ is calculated from the hard-sphere volumes of the Au atoms and remains constant throughout elongation.

\section{Hardware/Software Details}

Host: 2 x Quad-core Intel Xeon E5-2643, 3.3 GHz, 16 GB DDR3 RAM, PCI Express 3.0; Device: 4 x GeForce GTX 680 (all HOOMD-Blue simulations are performed on a single GPU), 1006 MHz, 1536 cores, 2 GB 256-bit GDDR5 RAM, PCI Express 3.0 x16; OS: CentOS 6.1 (64 bit). Software: HOOMD-Blue version 0.9.2, extended to include TB-SMA (written in CUDA in single-precision floating-point format; for more code details please refer to the Supporting Information), compiled with GCC 4.1.2, OpenMPI 1.4.5, and NVCC 5.0; LAMMPS version 21 March 2012, extended to include TB-SMA, compiled with GCC 4.1.2 and OpenMPI 1.4.5.

\section{Results and Discussion}

\subsection{TB-SMA benchmarks}

We first evaluate the performance of the TB-SMA potential in HOOMD-Blue. For comparison, we perform CPU-based simulations in LAMMPS \cite{Plimpton:1995}, which we have extended to include the TB-SMA potential. The LAMMPS simulations are performed in parallel across 8 cores; the HOOMD-Blue simulations are performed on a single GPU, with all other GPUs in the system idle. For consistency, the LAMMPS benchmarks are carried out on the host architecture of the HOOMD-Blue benchmarks. We perform simulations of unstretched Au NWs for 400 ps at 10 and 298 K, in all cases confirming that the CPU and GPU implementations of TB-SMA yields good agreement in the total energy and force magnitude at equilibrium (see Figure S2 in Supporting Information). We use a neighbor list buffer radius, $r_{buff}$, of 0.20 and 0.29 {\AA} in LAMMPS and HOOMD-Blue, respectively. In LAMMPS we obtain the best performance by decomposing the simulation box into equally sized domains along the [100] axis. We find the best HOOMD-Blue performance by turning off its particle sorting algorithm and ordering atoms along the [100]-axis.

%
%
\begin{figure}[h!]
       \centering
	\includegraphics[width=4.5in]{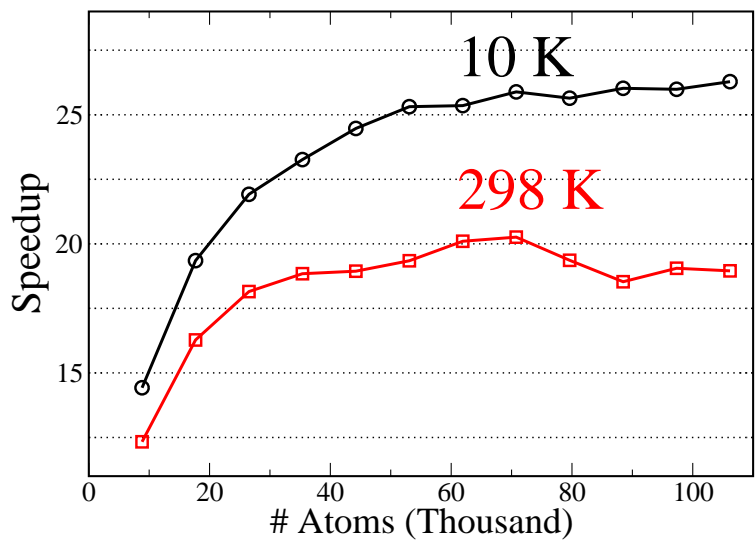}
	\caption{Performance speedup of TB-SMA in HOOMD-Blue running on a single GPU relative to the LAMMPS implementation run on 8 CPU cores. Speedup is the ratio of the timesteps completed per second on a GPU to that on the CPU implementation.
	\label{fig:speedup} }
\end{figure}

Figure 2 shows the speedup for the HOOMD-Blue simulations relative to the LAMMPS simulations at 10 and 298 K. The TB-SMA GPU implementation yields speedups between 12 and 27, depending on the temperature. The improved performance at low temperature (10 K) results from reduced atomic motion, which decreases the total number of neighbor list re-builds. We pause to emphasize here that the results in Figure 2 are specific to the compute environment we employ. We have performed additional benchmarks in high-performance computing environments employing different CPU, GPU, and compiler types, and while the speedups vary as compared to Figure 2, in all cases the GPUs examined provide increased performance relative to the CPU benchmarks, even for older generation GPUs (please refer to Supporting Information).     

Nevertheless, Figure 2 strikingly illustrates the types of performance gains possible for studies performed on a high-end desktop machine. The outstanding performance gain enables the study of a large number of independent trajectories for mapping out the landscape of Au NW failure behavior. For example, Figure 2 indicates that in the time required to run 10 replicates of a 100,000-atom NW at 10 K on eight CPU cores, roughly 260 replicates could be run on a single GTX 680. The ability to rapidly simulate NW elongation makes large-scale statistical studies more feasible.

In Table 2 we also show the absolute performance of the HOOMD-Blue simulations, which may be useful for readers who would like to compare their code performance to our own.   

%
%
\begin{table}[] 
\caption{Absolute performance (in timesteps per second) of GPU benchmarks. \label{table:runtimes}}
\begin{tabular}{c c c}
\hline
\hline
\# atoms & 10 K & 298 K \\
\hline
8823 & 2536 & 2286  \\
17673 & 1723 & 1507 \\
26523 & 1311 & 1141 \\
35373 & 1045 & 897 \\
44223 & 881 & 717 \\
53073 & 758 & 616 \\
61923 & 655 & 550 \\
70773 & 584 & 484 \\
79623 & 514 & 409 \\
88473 & 470 & 354 \\
97323 & 427 & 330 \\
106173 & 396 & 300 \\
\hline
\hline
\end{tabular}
\end{table}

\subsection{Variance in failure behavior for replicate simulations}

We next apply our GPU implementation of TB-SMA to study the variance in failure behavior of elongating Au NWs. All production runs are carried out on the Keeneland Initial Delivery System \cite{Vetter:2011} featuring NVIDIA Tesla M2090 GPUs (system specifications can be found in Supporting Information). Figure 3 shows a typical stress-strain curve for a NW with $D_{0}$ = 3.1 nm and $L_{0}$ = 20.4 nm elongated at 10 K. The serrations in the stress-strain curve indicate discreet dislocation events characteristic of ductile failure. Wu $et$ $al.$ \cite{Wu:2012} defined a critical NW length, $L_{C}$, that predicts the mode of NW failure; initial NW lengths exceeding $L_{C}$ undergo brittle failure, while initial lengths less than $L_{C}$ result in ductile failure. $L_{C}$ is given by

\begin{equation}
L_{C} = \frac{D_{0}}{\epsilon_{y}}cot(\alpha),\label{critical-length}
\end{equation} where $\epsilon_{y}$ represents the yield strain and $\alpha$ is the angle between the direction of dislocation slipping and the direction of the tensile load. Han and co-workers \cite{Han:2012} noted that $\alpha$ depends on whether deformation occurs via partial or perfect dislocation(s). For a [100]-oriented nanowire, $cot(\alpha)$ ranges from 1 for perfect-dislocation-mediated deformation to $\sqrt{2}$ for partial-dislocation-mediated deformation. Applying these values to eq 6 for a NW with $D_{0}$ = 3.1 nm and $\epsilon_{y}$ = 0.076, $L_{C}$ falls between 40.8-57.7 nm. Thus, the NW in Figure 3, whose initial length of 20.4 nm is well below $L_{C}$, undergoes the failure mode (ductile) predicted by eq 6.       

%
%
\begin{figure}[]
       \centering
	\includegraphics[width=4.0in]{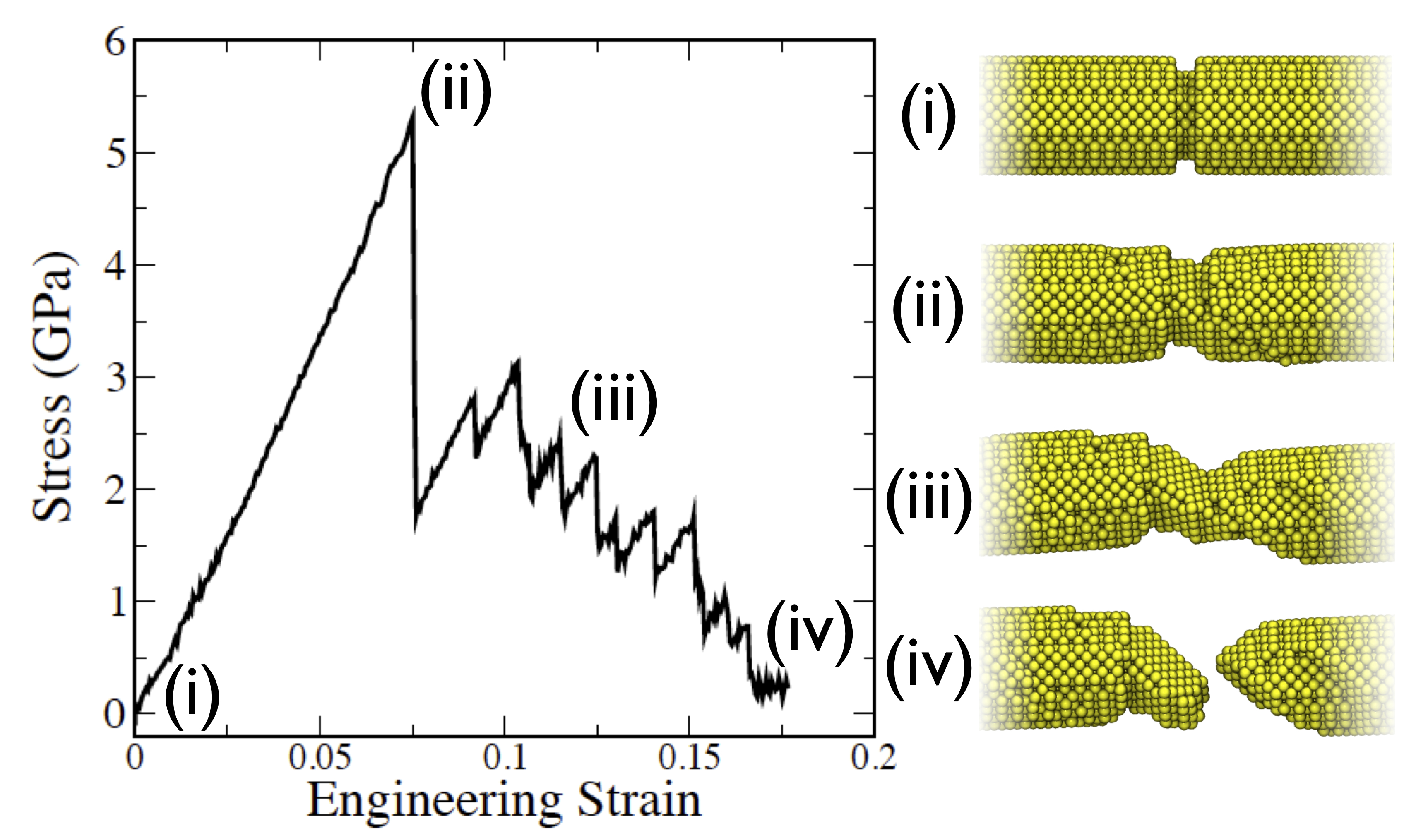}
	\caption{Stress-strain curve of an elongating Au NW ($D_{0}$ = 3.1 nm, $L_{0}$ = 20.4 nm), with zoomed-in images of the NW neck at select points.
	\label{fig:typical}}
\end{figure}

%
%
\begin{figure}[]
       \centering
	\includegraphics[width=4.0in]{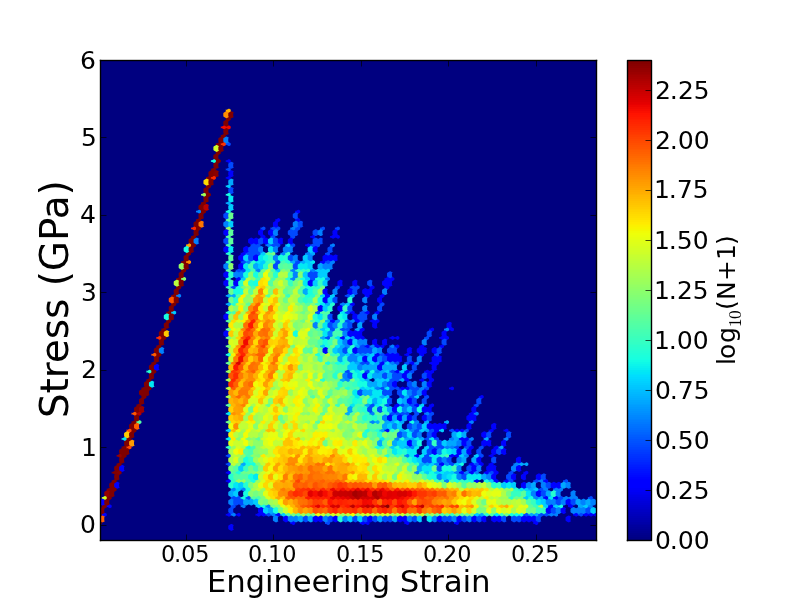}
	\caption{Stress-strain heat map constructed from 380 independent simulations of a NW with $D_{0}$ = 3.1 nm and $L_{0}$ = 20.4 nm elongated at 10 K. 
	\label{fig:heatmap}}
\end{figure}

%
%
\begin{figure}[]
       \centering
	\includegraphics[width=4.0in]{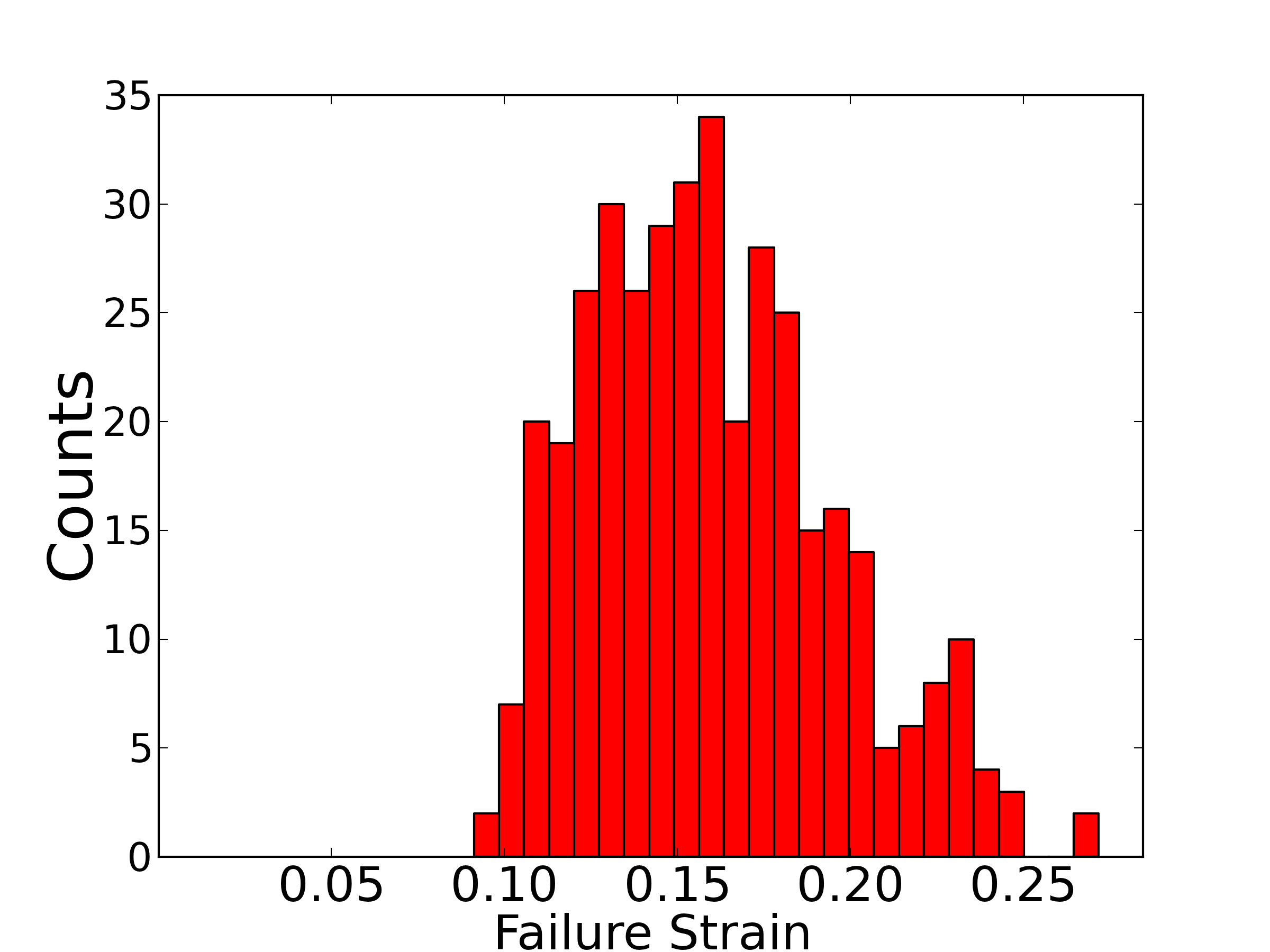}
	\caption{Histogram of the failure strain from 380 independent simulations of a NW with $D_{0}$ = 3.1 nm and $L_{0}$ = 20.4 nm elongated at 10 K. 
	\label{fig:failure-strain}}
\end{figure}

To investigate the role of stochastic events on the rupture process, we perform a total of 380 simulations like the one in Figure 3, with the only difference between replicates being the initial atomic velocity distribution. Figure 4 plots the stress-strain relationship for these 380 runs as a heat map, with bright areas corresponding to frequently occurring pathways. Prior to the yield point, the stress-strain pathway is very consistent between runs; however, following the yield point a region of diverse behavior emerges. This region is characterized by brightly colored diagonal streaks, which represent common stress-strain pathways and indicate the presence of plasticity. The streaks are faint at high values of strain as it becomes less likely for independent pathways to coincide. Finally, a bright horizontal area appears between strain values of $\sim$0.12-0.19 where many of the NWs have ruptured and exhibit a small residual stress following failure (note that stress-strain data is collected for 2 \AA\ of elongation following failure). In Figure 5 we plot the histogram of failure strains, confirming that many of the NWs fail in the $\sim$0.12-0.19 range. The wide range of failure strains in Figure 5 is surprising given the extremely low temperature at which the NWs are stretched, demonstrating the strong sensitivity of dislocation formation and behavior to the variance in relative atomic positions arising from stochastic thermal fluctuation.

%
%
\begin{figure}[h!]
       \centering
	\includegraphics[width=4.0in]{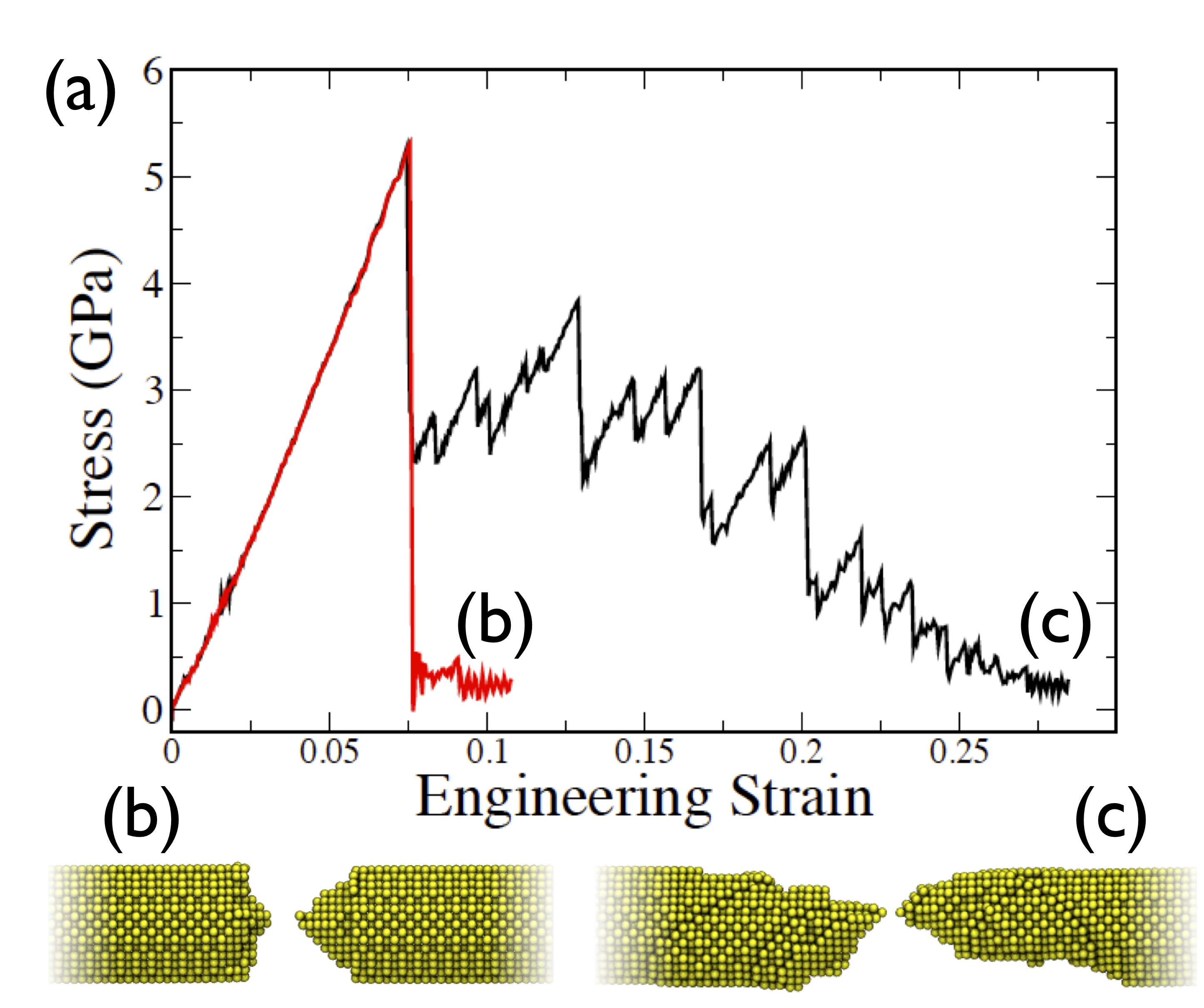}
	\caption{(a) Stress-strain curves for replicate runs of a Au NW ($D_{0}$ = 3.1 nm, $L_{0}$ = 20.4 nm) elongated at 10 K. In one case the NW undergoes (b) brittle failure while in another the wire undergoes (c) ductile failure. Zoomed-in snapshots immediately after NW failure are shown below. }
	\label{fig:brittle-vs-ductile}
\end{figure}

Figure 4 suggests that a vast majority of the 380 runs undergo ductile failure, in accordance with the predicted behavior from eq 6. However, there are a small number of cases in which the NW exhibits stress-strain behavior and post-rupture structure characteristic of brittle failure. This is illustrated in Figure 6a, where stress-strain data is plotted for the runs resulting in the lowest and highest failure strains. The red curve in Figure 6a drops off quickly following the yield point, and then remains relatively flat until the NW eventually fails. The lack of serrations in the stress-strain curve suggests that the NW experiences  little plastic deformation during elongation. The snapshot of the rupture region of the NW in Figure 6b also shows evidence of shearing along a single plane and no necking. In contrast, the black curve in Figure 6a exhibits numerous stress-strain serrations while the snapshot in Figure 6c shows evidence of significant slipping and necking. This result indicates that for this small-diameter NW elongated at 10 K, stochastic events are prominent enough to occasionally overcome rupture mechanisms dictated by NW size.

\subsection{Role of temperature}

To investigate the role of temperature in NW failure, we perform additional sets of simulations for a NW with $D_{0}$ = 3.1 nm and $L_{0}$ = 40.6 nm. Note that this length is just below the transition value ($L_{C}$ = 43.0 nm) predicted by the ductile-to-brittle model (eq 6) for partial-dislocation mediated deformation. Two hundred independent simulations are performed for each of four temperatures: 10, 100, 200, and 298 K. We note that this temperature range encompasses values applied in experiment \cite{Tsutsui_nl:2008,Yanson:1998,Scheer:1998}, and is well below the melting point of small Au NWs \cite{Pu:2007}. 

\begin{figure}[]
       \centering
	\includegraphics[width=6.0in]{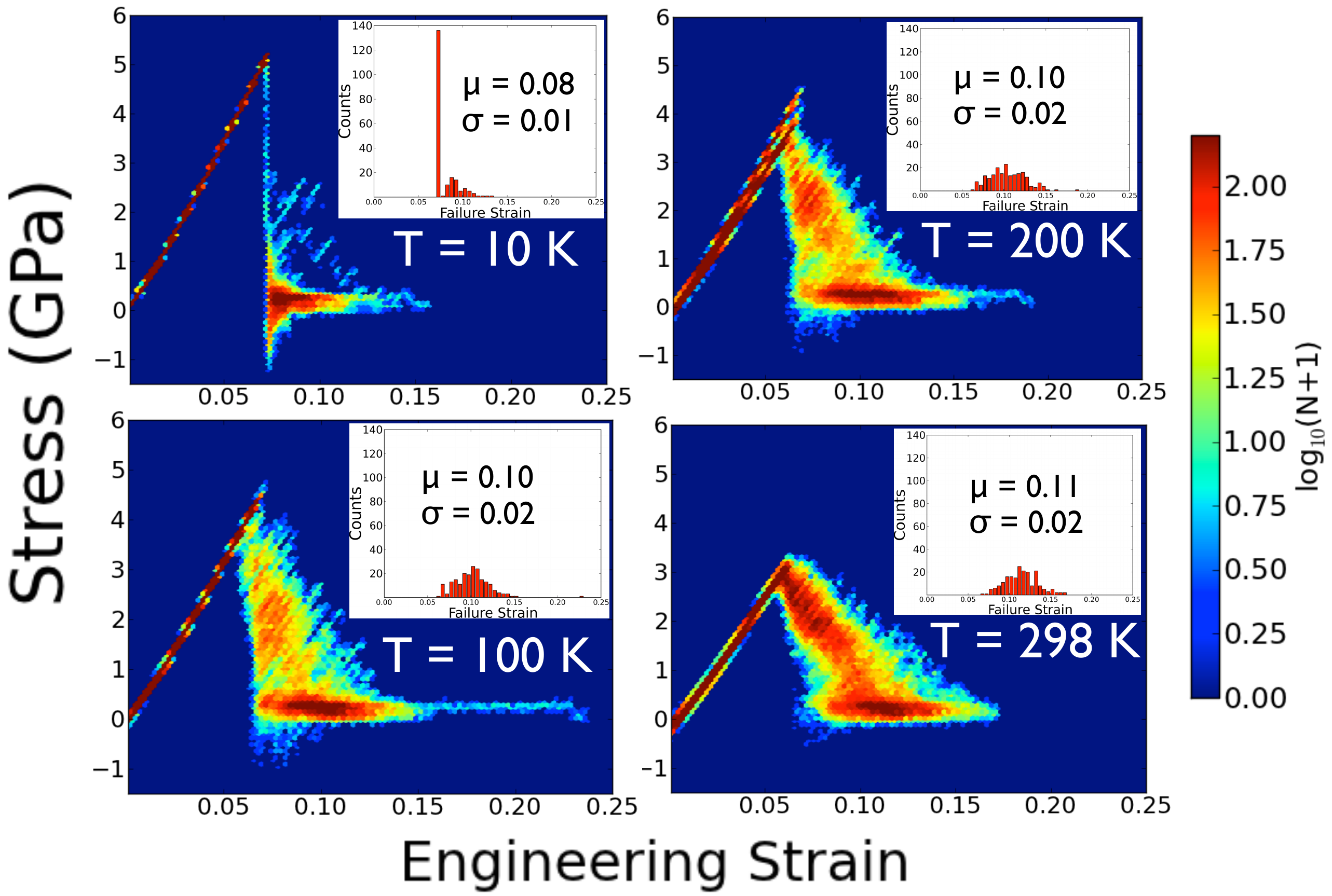}
	\caption{Stress-strain heat maps for a Au NW with $D_{0}$ = 3.1 nm and $L_{0}$ = 40.6 nm at four different temperatures. Two hundred independent simulations are performed at each temperature. The failure strain histograms, along with their corresponding average ($\mu$) and standard deviation ($\sigma$), are inset.
	\label{fig:temperature-plots}}
\end{figure}

Distinct temperature-dependent behavior is apparent from Figure 7. Prominent brittle failure can be observed in the heat maps by the presence of bright spots close to zero stress immediately after the yield point, whereas ductile failure exhibits brightly colored serrations extending well beyond the yield point. In Figure 7 the NWs fail in a predominantly brittle manner at 10 K and become significantly more ductile as the temperature is increased. Enhanced ductility and plasticity have been reported previously \cite{Koh:2005} for NWs elongated at higher temperatures, owing to the increased magnitude of atomic oscillations about the atoms' equilibrium positions, which promotes reconstruction of the crystal lattice. This effect decreases the yield strain, $\epsilon_{y}$, and yield stress, $\sigma_{y}$, of the NW at higher temperatures, effectively reducing the amount of energy available to drive deformation. The result of this is a tendency towards higher ductility and a reduction in the size and slope of the stress-strain serrations at high temperature, as seen in Figure 7. 

At 10 K, the NW experiences predominantly brittle failure, with a majority of the NWs failing immediately after yielding. This behavior is striking in comparison to Figures 4 and 5, where a NW with half the initial length exhibits predominantly ductile failure behavior. The NW with $D_{0}$ = 3.1 nm undergoes a clear ductile-to-brittle transition when $L_{0}$ is increased from 20.4 to 40.6 nm. Eq 6, which predicts $L_{C}$ between 40.8-57.7 nm, appears to slightly overpredict $L_{C}$ for NWs with $D_{0}$ = 3.1 nm elongated at 10 K. 

\subsection{Size-dependent failure behavior}

We next perform sets of simulations at the two extreme values of temperature (10 and 298 K) for six different NW sizes. The initial NW diameter is fixed at 3.1 nm while the length is varied from 20.4 to 121.3 nm. 100 independent simulations are run for each NW size and temperature, except for the NW with $L_{0}$ = 121.3 nm, where 93 runs are performed at 10 K and 88 runs are performed at 298 K. As we will show, the failure behavior for the NW with $L_{0}$ = 121.3 nm is highly reproducible so fewer replicates are needed.

\begin{figure}[]
       \centering
	\includegraphics[width=4.0in]{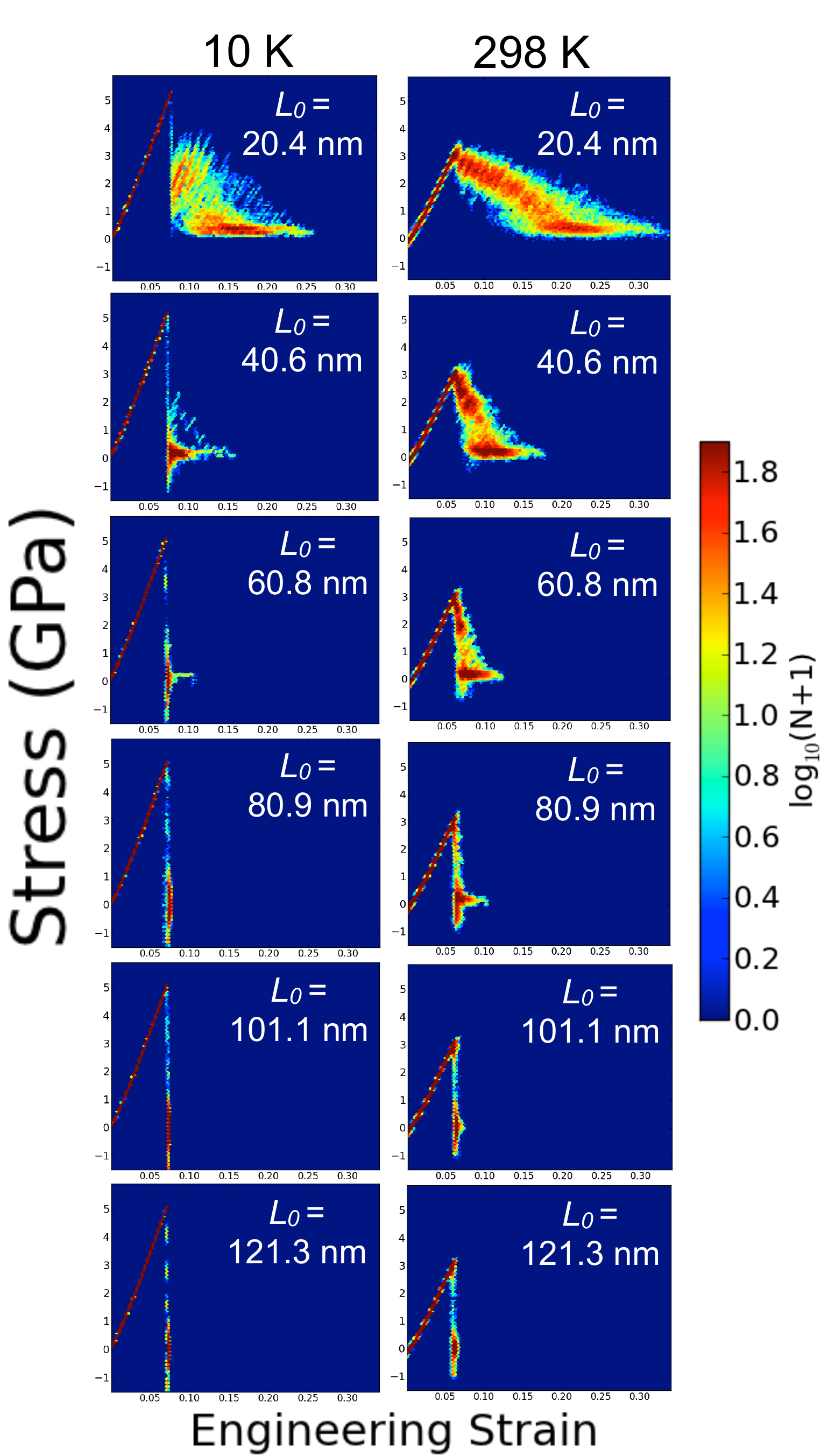}
	\caption{Stress-strain heatmaps for NWs with $D_{0}$ = 3.1 nm and varying lengths. The left column corresponds to simulations run at 10 K while the right column shows results at 298 K.  
	\label{fig:size-plots}}
\end{figure}

\begin{figure}[]
       \centering
	\includegraphics[width=4.0in]{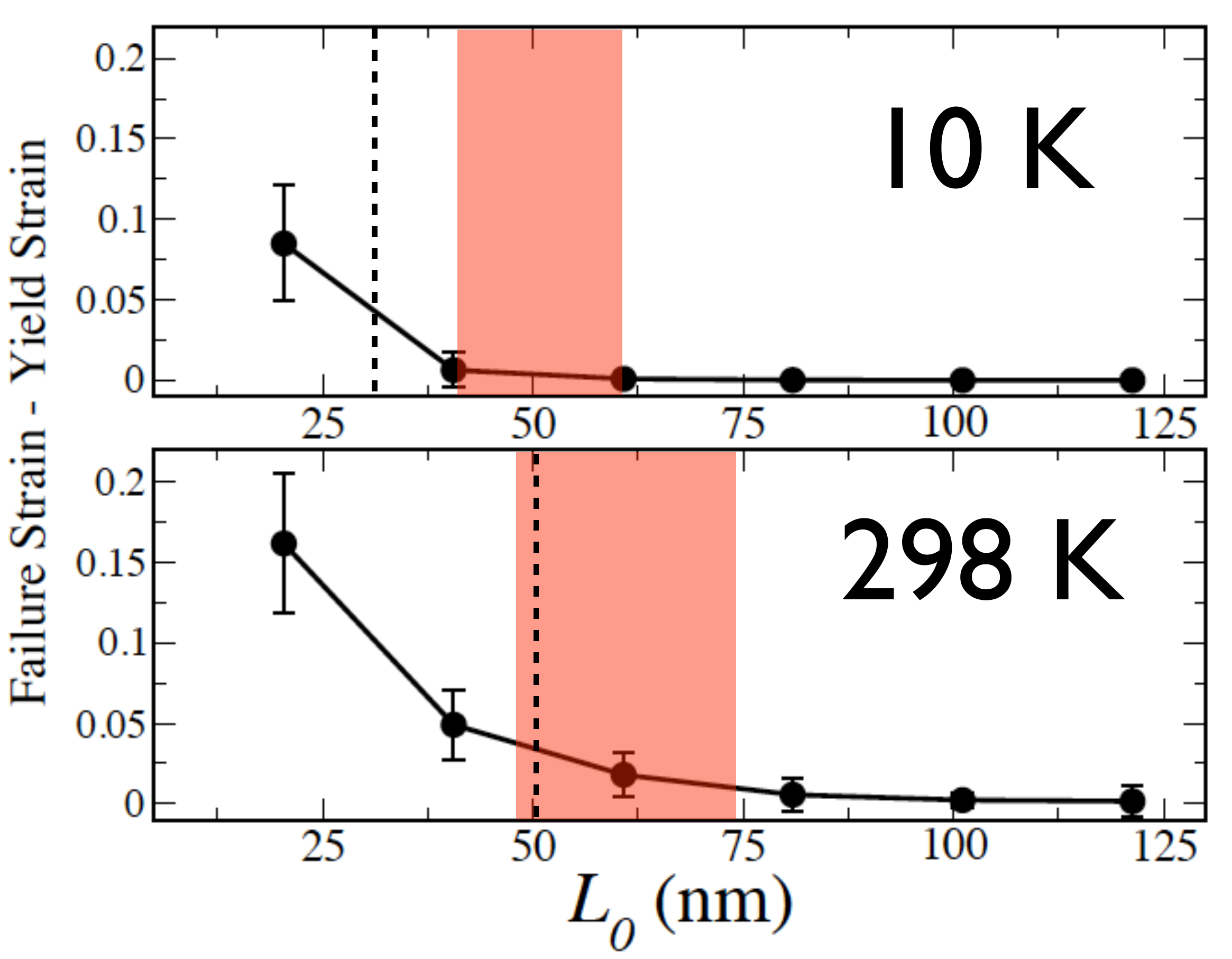}
	\caption{Strain after yielding for NWs with $D_{0}$ = 3.1 nm as a function of initial NW length at (top) 10 K and (bottom) 298 K. The dashed lines separate the ductile (left of the dashed line) and brittle (right of the dashed line) failure regions, as indicated by our simulation results. The colored region corresponds to the range of $L_{C}$ values predicted by eq 6.  
	\label{fig:failure-minus-yield-strain}}
\end{figure}

Figure 8 shows a clear transition from ductile to brittle failure with increasing $L_{0}$ at both temperatures. Serrations are present at lower values of $L_{0}$ but disappear at larger lengths. The transition occurs at a higher value of $L_{0}$ at 298 K compared to 10 K due to the aforementioned enhanced ductility effect. This behavior is also predicted by eq 6, with higher temperatures resulting in lower yield strain values. An instructive metric for quantifying failure mode is the total amount of strain that occurs after NW yielding. NWs that fail catastrophically feature very little strain following the yield point, while NWs undergoing plastic deformation are able to withstand some degree of strain after yielding. Figure 9 plots the strain after NW yielding as a function of $L_{0}$ for the two different temperatures. At 10 K, the strain after yielding is relatively high for the smallest value of $L_{0}$, but drops off quickly at $L_{0}$ = 40.6 nm. These data correspond to the NWs discussed previously (see Figures 3 and 7), and agree well with our previous interpretation that for $L_{0}$ = 20.4 nm the NW fails by a predominantly ductile mode while at $L_{0}$ = 40.6 nm a brittle mechanism is dominant. At initial lengths exceeding 40.6 nm the strain after yielding is minimal, indicating that the NWs are failing in a brittle manner. The exception to this is at $L_{0}$ = 60.8 nm, where there is evidence of occasional plasticity in the stress-strain heat map. The small error bars at high initial NW lengths also demonstrate the decreased variability in failure behavior within the brittle regime. At 298 K, the NWs experience extensive plasticity and exhibit high ductility at $L_{0}$ $<$ 60.8 nm. The strain after yielding drops to a small value of $\sim$0.01 at $L_{0}$ = 60.8 nm, where brittle failure is the prominent rupture mode. Occasional plasticity is observed at $L_{0}$ = 80.9 nm and $L_{0}$ = 101.1 nm before exclusively brittle behavior occurs at $L_{0}$ = 121.3 nm.

\begin{table} [h!]
\caption{Summary of mechanical properties for Au NWs with $D_{0}$ = 3.1 nm. Standard deviation is reported only in cases where it exceeds 10\% of the average value. \label{table:mechanical-summary} }
\begin{tabular}{ c  c  c  c  c  c  c  c  c}
    \hline \hline 
$L_{0}$ & \multicolumn{2}{c}{$\sigma_{y}$ (GPa)} & \multicolumn{2}{c}{$\epsilon_{y}$} & \multicolumn{2}{c}{$E$ (GPa)} & \multicolumn{2}{c}{$\epsilon_{f}$} \\\cline{2-9}
(nm) & 10 K & 298 K & 10 K & 298 K & 10 K & 298 K & 10 K & 298 K \\
     \hline
20.4 & 5.29 & 3.2 & 0.076 & 0.063 & 69.6 & 54.4 & 0.16 $\pm$ 0.04 & 0.23 $\pm$ 0.04 \\
40.6 & 5.18 & 3.1 & 0.073 & 0.061 & 71.6 & 54.8 & 0.08 $\pm$ 0.01 & 0.11 $\pm$ 0.02 \\
60.8 & 5.10 & 3.0 & 0.072 & 0.060 & 72.2 & 54.9 & 0.073 & 0.07 $\pm$ 0.01 \\
80.9 & 5.07 & 3.0 & 0.072 & 0.060 & 71.9 & 54.7 & 0.072 & 0.065 \\
101.1 & 5.13 & 3.0 & 0.072 & 0.059 & 72.3 & 54.8 & 0.072 & 0.062 \\
121.3 & 5.10 & 3.0 & 0.072 & 0.059 & 72.3 & 54.5 & 0.072 & 0.060 \\
\hline \hline
\end{tabular}
\end{table}

A summary of the average mechanical properties of the six NWs at 10 and 298 K is presented in Table 3. The yield stress, $\sigma_{y}$, yield strain, $\epsilon_{y}$, and Young's modulus, $E$, are higher at 10 K, and are not a strong function of NW length at either temperature.  The values for $\sigma_{y}$ and $E$ agree well with previously reported \cite{Park:2005,Han:2012,Koh:2006} values from Au NW simulations, and the strength of the Au NWs are significantly larger than bulk Au, in agreement with experimental results \cite{Seo:2011}. It is important to note that, even though $E$ is a measure of stiffness and clearly depends on temperature, it does not demonstrate a significant length-dependence and thus $E$ alone is not likely to be a meaningful predictor of the ductile-to-brittle transition. 

Also plotted in Figure 9 are the predicted $L_{C}$ ranges and the observed $L_{C}$ values from our simulations. The observed $L_{C}$ value is taken as the midpoint between the largest $L_{0}$ exhibiting predominantly ductile behavior and the smallest $L_{0}$ exhibiting predominantly brittle failure. As discussed previously, the NW with $D_{0}$ = 3.1 nm undergoes a ductile-to-brittle transition below the predicted $L_{C}$ range. On the other hand, at 298 K the observed $L_{C}$ value falls in the range predicted by eq 6. The observed $L_{C}$ value has some uncertainty ($\pm$ 10 nm) associated with it, as the length of the simulated NWs is changed in $\sim$20 nm increments. Nevertheless, eq 6 predicts $L_{C}$ with a fair amount of quantitative accuracy, and provides a reasonable initial guess for the true value of $L_{C}$. Eq 6 also accurately predicts the dependence of $L_{C}$ on temperature, as the range predicted by eq 6 shifts to larger values at 298 K, in agreement with our simulation results. Eq 6 is able to capture this because $\epsilon_{y}$ is a function of temperature; in other words, there is an implicit temperature-dependence built-in to eq 6. As noted previously, the failure behavior can exhibit some variability, especially close to $L_{C}$. From the NW sizes we simulate, the failure behavior is always a predominance ($\ge$95\%) of one failure mode (ductile or brittle). However, there may be characteristic NW sizes where an approximately even mixture of ductile and brittle failure occurs. 

There is a clear tendency towards ductile behavior with decreases in NW length, and likewise, an increased tendency to brittle failure as NW length is increased. According to eq 6, changes to the NW diameter should also affect failure behavior. To test this, we perform additional simulations at 10 K for two NWs, one with $D_{0}$ = 4.4 nm, $L_{0}$ = 20.5 nm and the other with $D_{0}$ = 6.0 nm, $L_{0}$ = 20.5 nm, to confirm the increased ductility effect for NWs with larger diameters. 204 independent simulations are performed for these NWs; in all simulations the NWs fail via a ductile mechanism (see Figure 10), with the NWs becoming more ductile with increases in $D_{0}$. Thus, the NW aspect ratio ($L_{0}$/$D_{0}$) is a critical parameter that may be used to adjust the degree of NW ductility. 

\begin{figure}[h!]
       \centering
	\includegraphics[width=3.2in]{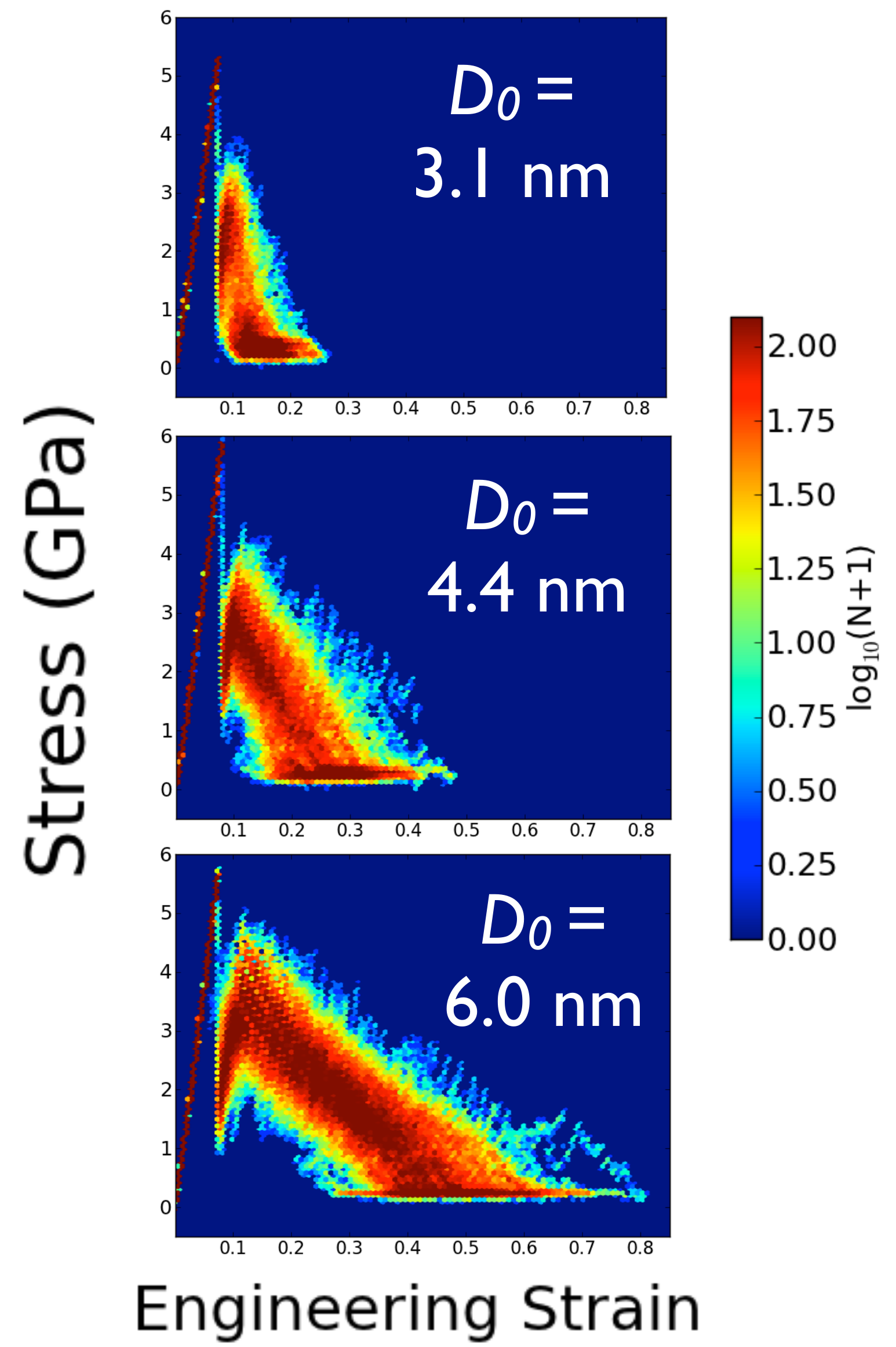}
	\caption{ Stress-strain heatmaps for Au NWs with $L_{0}$ = 20.4-20.5 nm and increasing (from top to bottom) diameter. 
	\label{fig:diameter-heatmaps}}
\end{figure}

It is surprising that the ductile-to-brittle model works well for the small-diameter NWs we consider here, as eq 6 was derived assuming the NWs obey classical dislocation theory. Prior work \cite{Iacovella:2011} has shown that the formation of noncrystalline structure ($e.g.$, polytetrahedra)  is promoted in small-diameter NWs elongated at high temperature. However, the formation of polytetrahedra was significantly reduced with increases in NW diameter from 1.1 to 1.9 nm. As the NWs in the present study are even larger, we expect limited noncrystalline structure formation, and thus significant departures from classical dislocation behavior are unlikely. Furthermore, it is possible that small-diameter, non-single-crystalline ($e.g.$, core-shell\cite{Oshima:2003,Gall:2004}) NWs are governed by different elongation mechanisms than the NWs simulted here. This possibility should be addressed in future studies. 

\section{Conclusions}

We have carried out a large number of simulations ($>$ 2000) to probe the validity and scope of the ductile-to-brittle transition in Au NWs ranging in length from $\sim$20-120 nm and diameter from $\sim$3-6 nm. We extended, and statistically confirmed, the applicability of the ductile-to-brittle transition to diameters as small as 3.1 nm, although $L_{C}$ is slightly over-predicted at low temperature (10 K). This was a somewhat surprising result since the ductile-to-brittle model was developed assuming classical dislocation theory applied. We therefore conclude that structures that may cause deviations from classical dislocation theory, such as polytetrahedra, do not form readily in these small-diameter wires. We further demonstrated that temperature plays an important role in the ductile-to-brittle transition, and can be used to tune failure behavior. The nanowire critical length was found to depend on temperature, as it was higher at 298 K than 10 K. Finally, stochastic events due to thermal fluctuation were found to be prominent enough to occasionally cause non-characteristic failure behavior based on the NW size; this was observed even at a low temperature of 10 K, where the effects of thermal motion should be minimal. These results provide comprehensive, statistical insight into NW failure that should be helpful for the controllable construction of nano- and atomic-scale devices. In particular we are intrigued by the possibility of using NW size to control NW tip structure in experiments of metal-molecule-metal junctions. This may prove a useful strategy for improving control over single-molecule conductance. 

Our study also adds to the growing body of literature that demonstrates the utility of GPU-based computing for high-throughput simulations studies requiring large-scale statistical analysis. This was enabled by porting the TB-SMA potential to HOOMD-Blue, an open-source MD package that runs on GPUs. Benchmarks of the TB-SMA code showed significant speedups for the single-GPU simulations relative a CPU implementation run across 8 CPU cores. We plan to contribute the TB-SMA code to the standard release of HOOMD-Blue so that future researchers can benefit from the performance gains of the code. 

\section{Acknowledgments}

WRF acknowledges support from the U.S. Department of Education for a Graduate Assistance in Areas of National Need (GAANN) Fellowship under grant number P200A090323, as well as the U.S. Department of Energy under grant number DEFG0203ER46096.  AKP received support from a Vanderbilt Undergraduate Summer Research Program (VUSRP) fellowship. APS received support through a Research Experiences for Undergraduates (REU) Fellowship from the National Science Foundation (NSF) under grant number DMR-1005023. CRI and PTC acknowledge support from NSF through grant CBET-1028374. This work was supported by computational resources at the National Institute for Computational Sciences, Project-ID UT-TNEDU014 \cite{Vetter:2011}, and also National Energy Research Scientific Computing Center, which is supported by the Office of Science of the U.S. Department of Energy under Contract No. DE-AC02-05CH11231.

\vspace{0.5in}

\noindent {\bf Supporting Information Available:} Additional benchmarks and GPU code details. This material is available free of charge via the Internet at http://pubs.acs.org. 

\bibliography{new-library}

\section{TOC Image}

\begin{figure}[h!]
       \centering
	\includegraphics[width=4.0in]{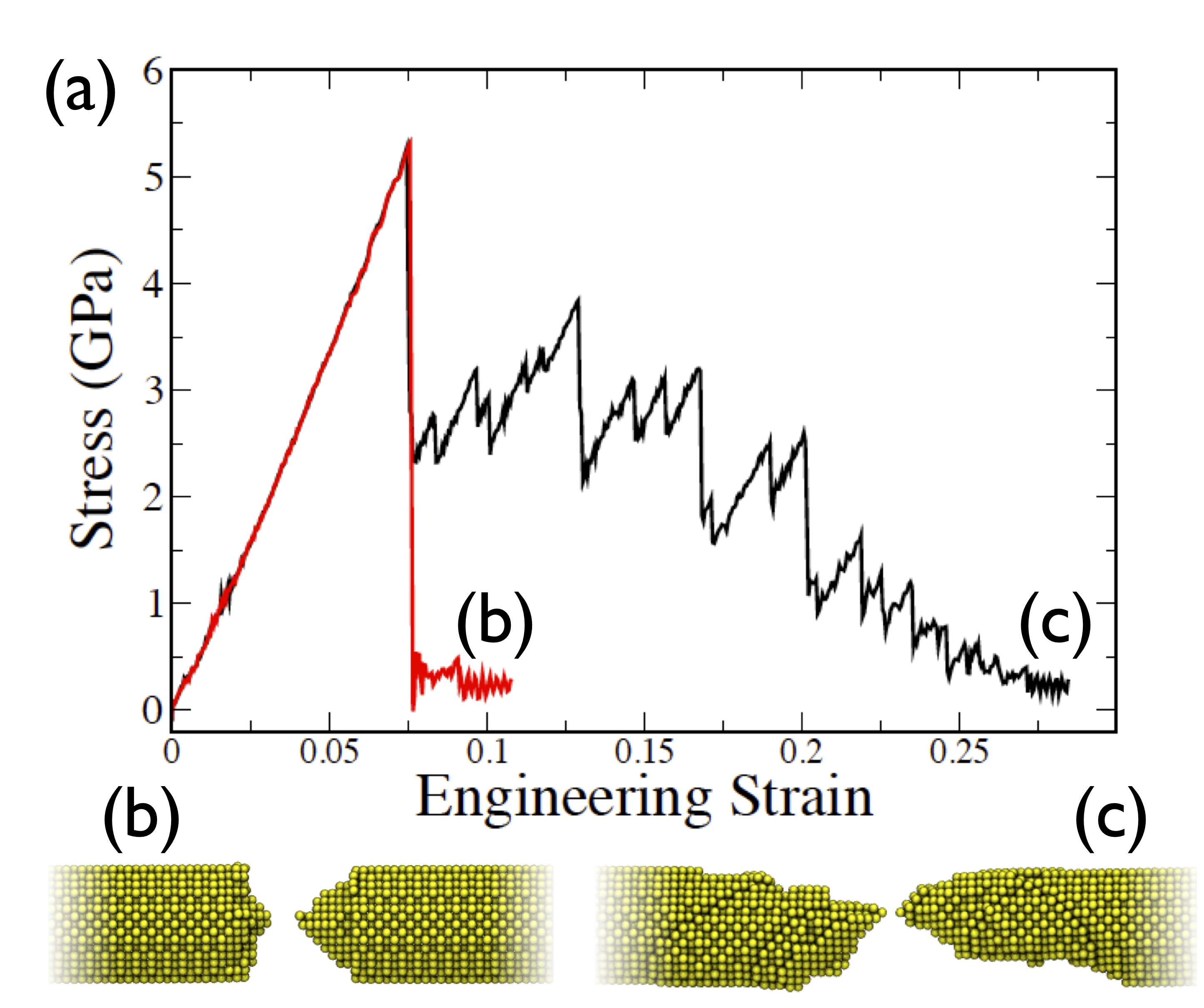}
\end{figure}

\end{document}